# A Visual Cryptographic Encryption Technique for Securing Medical Images

Quist-Aphetsi Kester, MIEEE

*Faculty of Informatics, Ghana Technology University College, PMB 100 Accra-Norths, Tesano, Accra, Ghana*

*Abstract*— The increased growth in the use of transmission of multimedia medical contents over unsecured and open networks provides insecurity for confidential patient information over these networks. Digital encryption of medical images before transmission and storage is proposed as a way to effectively provide protection of patient information. Encryption before watermarking of these images is necessary in order to ensure inaccessibility of information to unauthorized personnel with patient.

This paper presented a visual cryptographic technique for encrypting of medical images before transmission or storage of them. This will make such images inaccessible by unauthorized personnel and also ensures confidentiality. The process made use of an encryption technique that is based on pixel shuffling and a secret key generated from the image.

*Keywords*—visual cryptography, medical images, secret key, encryption

## I. INTRODUCTION

The usage of the internet for the transmission of multimedia content has become a very frequent medium for the exchange of digital information almost all institutions that are using the internet. It is therefore important to secure data over open and unsecured networks in order to ensure safety of sensitive data. Medical information of patients are sensitive and needed to be protect during storage, especially in the cloud, and during transmission between two hospitals. Hence the usage of cryptography in the protection of such data is very crucial.

In cryptography, encryption processes are used in transforming information using an algorithm to make it unreadable to anyone except those possessing special knowledge, usually referred to as a key. The result of the process is encrypted information. The reverse process is referred to as decryption. [1] Cryptography has evolved from the from classical such as Caesar, Vigenère, Trifid ciphers to modern day cipher and public key systems such as Diffie-Hellman etc[2]

The cryptography in digital computing has been applied to different kinds of digital file formats such as text, images video etc[3][4].One of the best-known techniques of visual cryptography has been credited to Moni Naor and Adi Shamir.

They demonstrated a visual secret sharing scheme, where an image was broken up into n shares so that only someone with all n shares could decrypt the image, while any n − 1 shares revealed no information about the original image. Each share was printed on a separate transparency, and decryption was performed by overlaying the shares. When all n shares were overlaid, the original image would appear [5].

Visual cryptography encodes a secret binary image into n shares of random binary patterns. The secret image can be visually decoded by superimposing a qualified subset of transparencies, but no secret information can be obtained from the superposition of a forbidden subset [6]. By engaging a cryptographic encryption technique involving pixel shuffling and inter changing their position to create the ciphered image, this proposed method makes it difficult for decryption of the image without prior knowledge of the algorithm and the secret key used. Secret shared key and visual cryptography are two distinct types of cryptography. In this paper a method is proposed which combines visual cryptography with shared secret key for the encryption and the decryption process.

This paper has the following structure: section II is about related works, section III is on the methodology employed for the encryption and the decryption process of the digital images, section IV presents the mathematical algorithms employed to come out with a ciphered image for the encryption process. Section V Results and analysis of the ciphered mages obtained from the implementation of the algorithm used in the encryption process, and section VI concluded the paper.

## II. RELATED WORKS

Security has become an inseparable issue not only in the fields strictly related to secure communications but fields that have anything to do with storage of data as well. Visual Cryptography is the study of mathematical techniques related aspects of Information Security which allows Visual information to be encrypted in such a way that their decryption can be performed by the human visual system, without any complex cryptographic algorithms. [7][8]

Mandal, J.K. and Ghatak, S., In proposed a novel (2, m + 1) visual cryptographic technique, where m number of secret images were encrypted based on a randomly generated master as a common share for all secrets which was decodable with any of the shares in conjunction with master share out of m + 1 generated shares. Instead of generating new pixels for share except the master share, hamming weight of the blocks of the secret images were been modified using random function to generate shares corresponding to the secrets. At the end of their work, the proposed scheme was secure and very easy to implement like other existing techniques of visual cryptography. At the decoding end the secrets were revealed by stacking the master share on any one share corresponding to the secrets in any arbitrary order with proper alignment directly by human visual system where shares were printed on different transparencies which conforms the optimality of using shares. The aspect ratio and dimension of the secret images and the generated shares with respect to the source images remained constant during the process [9].

Encryption is used to disguise data by using a mathematical algorithm to make it unintelligible to unauthorized users. Providing such security for transmitted data is very important especially when the data is being transmitted and stored across open networks such as the Internet. Since, image data have special features such as bulk capacity, high redundancy and high correlation among pixels that imposes special requirements on any encryption technique. Gilani, S.A.N. and Bangash, M.A. proposed an extension to the block-based image encryption algorithm (BBIE) scheme that works in combination with Blowfish encryption algorithm [10]. Whereas BBIE was meant for 256-color bitmap images, the proposed technique also handles RGB color images and, for the cases studied, improved the security of digital images. In this enhanced technique, which they call the enhanced block based image encryption technique (EBBIE) the digital image was decomposed into blocks, then two consecutive operations - rotating each 3D true color image block by 90deg followed by flipping row-wise down - are performed to complicated the relationship between original and processed image. These rendered blocks were then scrambled to form a transformed confused image followed by Blowfish cryptosystem that finally encrypted the image with secret key.

Experimental results show that the correlation between adjacent pixels was decreased in all color components and entropy was increased for the cases studied [11].

Quist-Aphetsi Kester developed a cipher algorithm for image encryption of m*n size by shuffling the RGB pixel values. The algorithm ultimately makes it possible for encryption and decryption of the images based on the RGB pixel. The algorithm was implemented effectively without change in the image size and was no loss of image information after drryption[13].

In order to achieve secure access to color medical image, a symmetric color medical image encryption was proposed in by Abokhdair, N.O., Manaf, A.B.A. and Zamani, M. Their algorithm was base on combination of scrambling and confusion processes. 2D lower triangular map was used for scrambling the addresses of image pixels, and the proposed propeller algorithm was used to confuse the gray values of image pixels [14].

One of the main aims of application of image encryption is to be able to transmit secure information between two or more parties over communication channels such that no unauthorized user can obtain and understand the transmitted content. In this paper, a visual cryptographic technique based on pixel shuffling and a shared secret key was used for the encryption process. The key was computed from the image and then used in the encryption process.

III. METHODOLOGY

In this encryption process, an input image which was a plain image was operated on by a function to generate a secret key from it. The key was then used to encrypt the image by shuffling the pixels of the plain image based on an algorithm. The ciphered image was obtained at the end and it can either be stored or transmitted over a communication network. The received image was then operated on again by a function to obtain the key in order to decrypt the image. The processes are shown in figure 1 below.

In figure 1, PI is the plain image and CI is the ciphered image. Sk is the secret key used in the encryption and the decryption process of the image. Enc.Alg is the encryption algorithm and Dec.Alg is the decryption algorithm employed.

### IV. THE ALGORITHM

This document is template. We ask that authors follow.

1. Start
2. Import data from image and create an image graphics object by interpreting each element in a matrix.
3. Get the size of r as [c, p]
4. Get the Entropy of the plain Image
5. Get the mean of the plain Image
6. Compute the shared secret from the image
7. Iterate step 8 to 17 using secret key value
8. Extract the red component as 'r'
9. Extract the green component as 'g'
10. Extract the blue component as 'b'
11. Let r =Transpose of r
12. Let g =Transpose of g
13. Let b =Transpose of b
14. Reshape r into (r, c, p)
15. Reshape g into (g, c, and p)
16. Reshape b into (b, c, and p)
17. Concatenate the arrays r, g, b into the same dimension of 'r' or 'g' or 'b' of the original image.
18. Finally the data will be converted into an image format to get the encrypted image.

The inverse of the algorithm will decrypt the encrypted image back into the plain image.

The secret key is obtained as follows:

$$Sk = [(c \times p) + |(He \times 10^3)| + |(\bar{x} = \frac{1}{n} \cdot \sum_{i=1}^{n} x_i)|] \mod p$$

Where c, p are dimension of the image and He is the entropy value of the image and *x* bar is the arithmetic mean for all the pixels in the image.

The implantation of the algorithm was done using MATLAB Version 7.12.0 (R2011a. The image sizes used were not fixed since the algorithm can work on mxn image size. The algorithm was written in m-file and tested on sample of medical images. The images were encrypted and the results were analyzed below.

### V. RESULTS AND ANALYSIS

Three sample of medical images were encrypted by the algorithm using MATLAB and the results are below. The RGB graphs from figure 5 to 9 were plotted using the first 10000 pixel values of both plain and ciphered images.

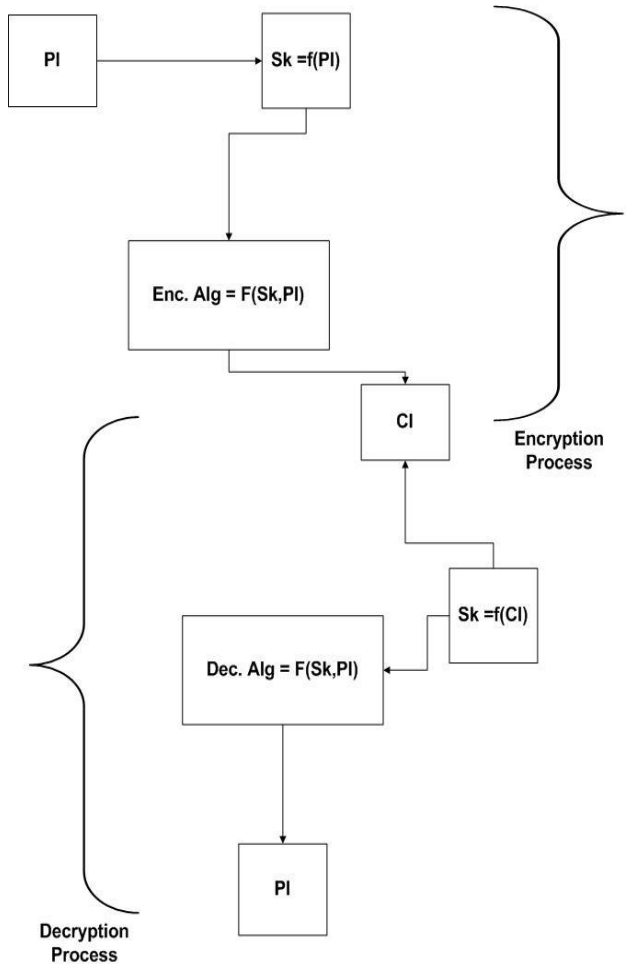

**Figure 1: The encryption and the decryption process**

In the encryption process, the images used had their RGB colours shuffled to obtain ciphered images. The ciphering of the images for this research was dependent solely on the RBG pixel values of the images and the secret key obtained from the image. There were no changes of the bit values of the images used and there was no pixel expansion at the end of the encryption and the decryption process. The numerical values of the pixels were displaced from their respective positions and the RGB values were interchanged in order to obtain the ciphered images. This implies that, the total change in the sum of all values in the image was zero. Therefore there was no change in the total size of the image during encryption and decryption process. The characteristic sizes of image remained unchanged during the encryption process.

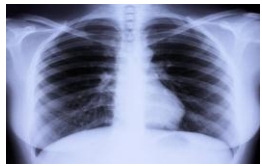 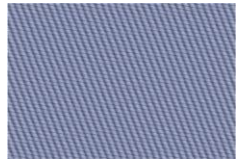

(a)Plain image [15]   (b) Ciphered Image
**Figure 2: X-ray picture of the ribs**

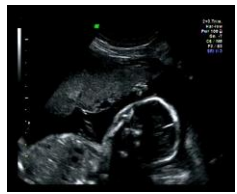 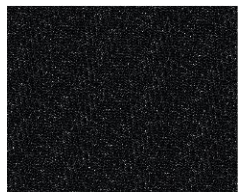

(a)Plain image [16]   (b) Ciphered Image
**Figure 3: Ultra sonic Image of the womb**

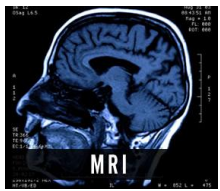 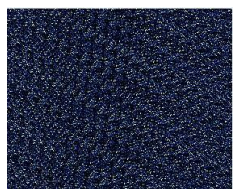

(a)Plain image [15]   (b) Ciphered Image
**Figure 4: Magnetic Resonance Imaging of the brain**

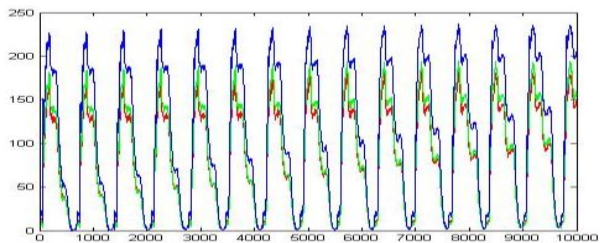

**Figure 5: An RGB graph of figure 2(a)**

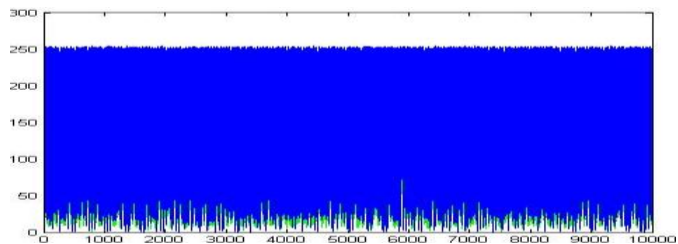

**Figure 5: An RGB graph of figure 2(b)**

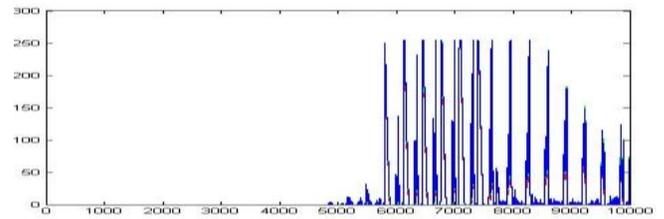

**Figure 6: An RGB graph of figure 3(a)**

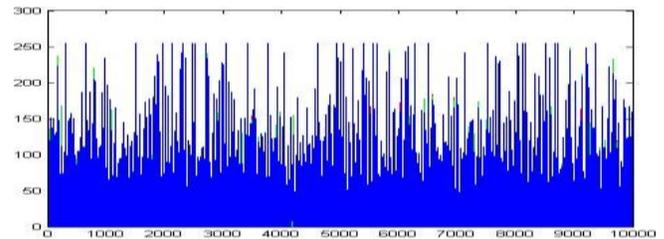

**Figure 7: An RGB graph of figure 3(b)**

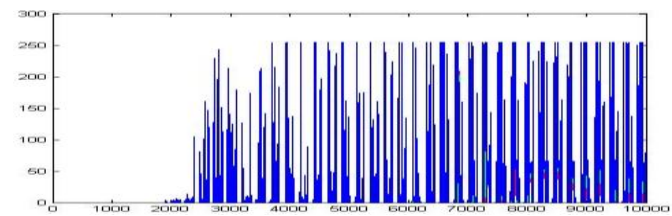

**Figure 8: An RGB graph of figure 4(a)**

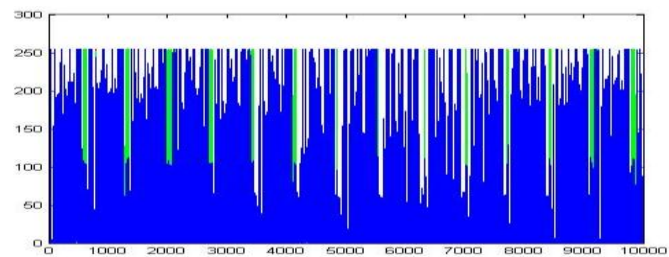

**Figure 9: An RGB graph of figure 4(b)**

## VI. CONCLUSION

The encryption process was effective for all the images and there was no pixel expansion at the end of the process. The entropy, mean and Secret key values for the images in figure 2, figure 3 and figure 4 were [7.6035, 143.5284, 135], [3.7603, 18.5810, 178], and [4.6894, 39.0415, 248] respectively. A slight change in the ciphered image size and pixel values resulted in a change in the decryption result.

This makes the algorithm very effective for closely related images. The total entropy and the mean of the plain images never changed for all the ciphered images and the plain images. That is the average total pixel before encryption was the same as the average total pixel after encryption.